# Plasmon drag and photoinduced magnetic effects in plasmonic and magnetic metals

Md Afzalur Rab, Terence Baker, and Natalia Noginova
*Department of Physics, Norfolk State University, Norfolk, VA 23504, USA*

**ABSTRACT.** Photoinduced electric effects in plasmonic and magnetic materials under pulsed laser illumination exceed the prediction of the electromagnetic momentum-transfer mechanism by orders of magnitude. In order to get more information on the nature of the effect we study the kinetics of photovoltages in permalloy thin films at different photoexcitation configurations. The photoinduced electric signals consist of magnetically dependent and magnetically independent components which are of significant magnitude and mainly follow the temporal profile of the laser pulse. In contrast, the predicted contribution from the Anomalous Nernst Effect is much weaker and exhibits substantially slower kinetics. These observations suggest that ultrafast photoexcited hot electrons, rather than thermal mechanisms, play the dominant role in generating the observed photovoltage.

## I.INTRODUCTION

Photoinduced electric currents in plasmonic metal films and nanostructures demonstrate a giant enhancement under surface plasmon resonance conditions [1-17], exceeding off-resonance values by orders of magnitude. As a rule, the direction of plasmon-induced photocurrents corresponds to the drift of electrons in the direction of surface plasmon polariton (SPPs) propagation. This plasmon drag effect (PLDE) allows one to directly incorporate plasmonic elements into electronic circuits and monitor plasmonic excitations electrically; this presents interest for various applications, for example, plasmonic sensors with compact electrical detection [14,15].

The origin of PLDE is not fully clear. The magnitudes of the currents significantly exceed predictions of the photonic pressure model even with account for SPP- enhanced absorption [3,4]. The effect is commonly attributed to the electromagnetic momentum transfer from light to free electrons [18,19] and explained with “plasmonic pressure” mechanism using the electromagnetic hydrodynamic momentum loss approach [20,21]. The “plasmonic pressure” model correctly predicts the direction of the photoinduced currents in various structures, however, fails to describe their magnitudes without additional strong assumptions [21].

In the experiments, PLDE has been observed in different materials, such as silver, gold, platinum, copper, aluminum and permalloy [1,2,16,22,23]. The magnitude of the effect normalized to the absorbed energy is strongly different in different materials. In most cases, the magnitude drops with the increase in the wavelength while in some materials and structures it can show a broad maximum [16]. This behavior does not well correspond to a simple plasmonic pressure model [20,21] and needs an additional or an alternative mechanism for a proper description.

One of the alternative mechanisms to be considered is related to the inter-band or intra-band transitions. The photoinduced electric currents in solids, in particular, semiconductors often involve inter-or intra-band transitions [24-26]. According to the model presented in [27,28], light-induced electric currents in solids can be related to transitions selective for electrons with velocities parallel or antiparallel to the optical *k* vector due to the Doppler effect. For electrons traveling against the light propagation direction, the frequency of the wave is higher than that for electrons moving in the same direction as the optical *k* vector, thus leading to different probabilities of the excitation to upper bands for electrons with the parallel and antiparallel velocities [27, 28]**.** Such momentum-selective transitions can produce a non-zero total electric current with the polarity determined by the band structure and mobilities of electrons and holes in the participating bands. This consideration can be applied to metals as well [27, 28], predicting *dc* photoinduced currents for an oblique incidence of illumination.

The model [27, 28] does not specifically discuss plasmon resonances, which are of key significance for PLDE. Note that in the PLDE experiment, a giant enhancement of photoinduced electric effects takes place at the resonance conditions while off-resonance currents are much lower and sometimes of a opposite polarity. In addition, PLDE is also observed in infra-red range in gold and silver which is far from inter-band transitions.

Band structures in ferromagnetic materials depend on spin polarization, and studies of photoinduced electric effects in magnetic metals with plasmonic properties can provide more information on the mechanism behind high magnitudes of photoinduced electric currents in plasmon drag effect. This inspires us to closely look at photoinduced voltages in permalloy.

Permalloy (Py) is Iron-Nickel alloy, soft ferromagnetic material with a very high magnetic permeability of 100,000 and a low coercivity of 0.5-5Oe [29,30]. Plasmonic properties of permalloy are poor due to high optical loss, however, broad SPP resonances (with Q-factors of 3-7 depending on the wavelength) have been observed in Py gratings in the visible range [31]. Under pulse laser illumination, the photovoltage peaks at the SPP resonance conditions, and shows a clear dependence on magnetic field with a characteristic hysteresis [22,23]. The width of the hysteresis depends on the coercive field magnitude for a particular structure, clearly indicating the role of magnetization. In [23] photovoltages observed in structures of different geometry were discussed as a result of two contributions, a magnetically independent component (plasmon drag) and a magnetically dependent (or magnetization-dependent) component which has the opposite polarities in the opposite fields. This magnetization-dependent component was tentatively ascribed to Anomalous Nernst (ANE) Effect [32,33] and explained with nonhomogeneous heating of the film by laser light due to the small penetration depth (~ 13 nm). However, common ANE voltages observed in permalloy in various experiments are in the order of nanovolts or microvolts [33] while the effects observed in the pulse laser experiments [23] can reach millivolts or even tens of millivolt, which would require significant temperature gradients of several degrees across the film thickness (~30-40 nm) [22]. Another possible mechanism is the Inverse Spin Hall Effect (ISHE) [34, 35], which in the laser experiments can be associated with spin currents arising due to laser induced gradients in spin polarization.

More detailed study is needed to better understand the nature of the strong photovoltages in plasmonic and magnetic structures. In this work, we closely look at the kinetics of the photovoltages in permalloy films and structures at different directions of illuminations. We also perform COMSOL simulations for temperature gradients and ANE-related voltages, compare them with experimental kinetics and discuss possible mechanisms of the effects.

## II.EXPERIMENTAL

The sample schematics and setup are illustrated in the FIG. 1. The permalloy (Py) films and gratings with the thickness of about 30 nm are fabricated with dc sputtering technique, using a glass slide or a profile modulated polycarbonate substrate derived from a commercial Blu-ray disc following the procedure [36]. The films are prepared in the shape of long strips with the length of about 12-15 mm and the width of about 3 mm.

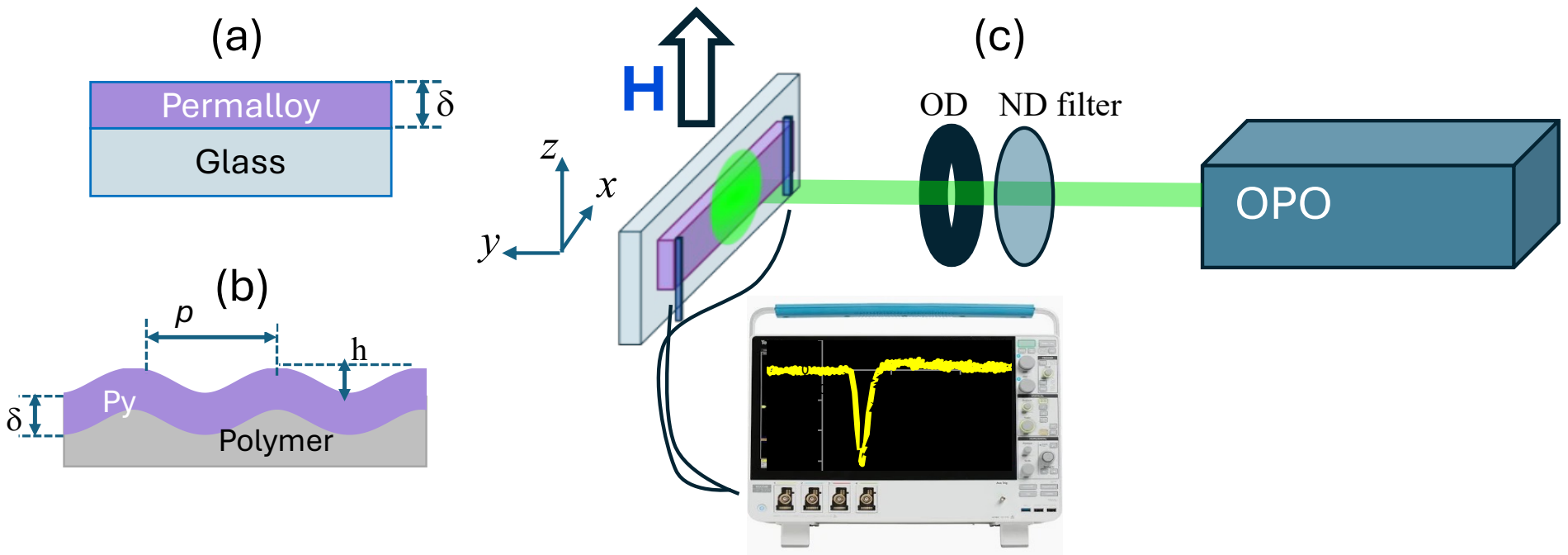

FIG. 1. Schematics of (a) Flat and (b) Blu-ray-based samples; (c) Experimental setup.

The electric contacts are attached with the conductive epoxy on the opposite ends. The electric resistance of the films is in the order of 300-600 Ohm. The laser light at p-polarization, pulse duration of 5 ns and wavelength tunable between 430-670 nm is supplied by the OPO laser. The pulse energy is controlled by the set of neutral density (ND) filters to achieve the energy incident on the sample of about 0.1-0.2 mJ. The illumination spot is centered in the middle of the sample strip; the spot diameter is regulated with optical diaphragm (OD) to be slightly higher than the film width (3 mm). The photoinduced electric signals are recorded with Tektronix digital oscilloscope (2 GHz, 50 Ohm internal resistance). The magnetic field with the magnitudes up to 90 Oe is supplied by an electromagnet.

The film on the glass substrate can be oriented in two ways, (i) Py film on the front side facing the laser and glass substrate on the back side or (ii) glass substrate on the front side and the film on the back side, thus allowing illumination of Py film from air or from glass. The goniometer stage is used to orient the Py/Blu-ray structure at various incidence angles.

Before presenting the experimental results, first, we calculate the expected temperature gradients and the corresponding ANE voltages using COMSOL simulations. The ANE effect is expected to be dominant in our geometry. (Although ordinary and planar Nernst effects can contribute to the voltage, but those contributions are expected to be negligibly weak in our geometry and magnetic fields.) These calculations establish the magnitude and temporal behavior expected from a purely thermoelectric mechanism and serve as a reference for interpreting the experimental photovoltage measurements discussed in the following sections. In our simulations, for simplicity, we consider flat films and two different configurations of the illumination, from air and from glass at the normal incidence. The gratings geometry and oblique illumination lead to a complex pattern of temperature gradients, which will be analyzed and discussed in detail elsewhere.

## III.NUMERICAL SIMULATIONS OF ELECTROMAGNETIC LOSSES AND TEMPERATURE GRADIENTS

Due to high optical loss in permalloy leading to a short penetration depth of the electromagnetic field, one can expect nonhomogeneous heating of the film by the laser pulse and generation of the voltage due to the Anomalous Nernst Effect [32,33]. The electric field in the presence of a thermal gradient $\nabla T$ and magnetization $M$, can be found as

$$E_{ANE} = -N\mu_0 M \times \nabla T, \quad (1)$$

where $N$ is the Nernst coefficient.

In order to estimate the temperature gradient and its kinetics in our films during and after the laser pulse, a two-step procedure is implemented using COMSOL software. COMSOL RF module and Heat Transfer module were used to estimate electromagnetic power loss density and temperature gradients respectively.

In the first step, electromagnetic loss, $q(y)$, in Py films is calculated for two illumination conditions: a) illumination from air, and b) illumination from glass, see FIG. 2 (a) and (b) correspondingly. We assume the wavelength of 600 nm, complex dielectric constants, $\varepsilon_r$ =-8.74 +19.1i in permalloy and 2.3 in glass [37]. In these simulations, we concentrate on flat films, however this procedure can be readily applied to any geometry including gratings. The loss profile for illumination from air is shown in FIG. 2(c). Illumination from glass gives a similar result.

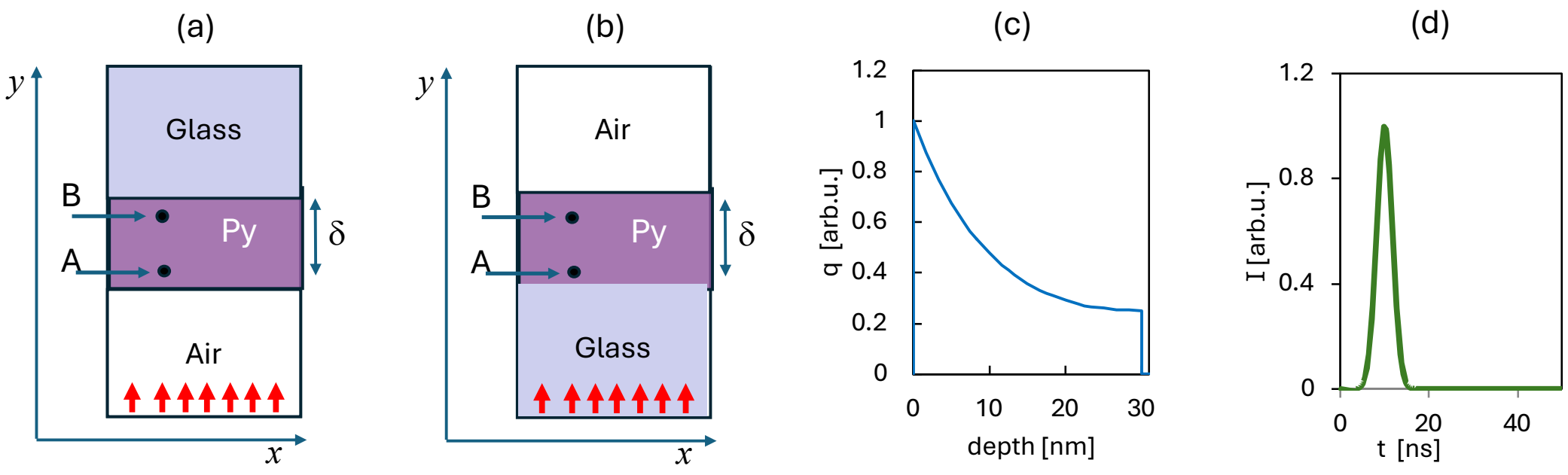

FIG. 2. Schematics of the model for illumination from (a) air and (b) glass; (c) loss profile for illumination from air; (d) pulse kinetics.

In the second step, we numerically solve the one-dimensional heat equation,

$$\frac{\partial u}{\partial t} = -\frac{\kappa}{cp}\frac{\partial^2}{\partial y^2} + \frac{1}{cp}Q(y,t), \quad (2)$$

where $u(y, t)$ is the difference between the local temperature at the particular depth and ambient temperature $T_0$, $\kappa$ is the thermal conductivity, $c$ is the heat capacity, $\rho$ is the density, and $Q(y, t)$ is the heat sources associated with electromagnetic loss. In the experiment, the diameter of the illuminated spot (3 mm) is much higher ($10^5$- fold) than the film thickness (~30 nm). Therefore, even if gradients exist across the spot width, the through-thickness gradient is expected to dominate. Thus, as a good approximation, we consider a one-dimensional case, where all functions depend only on one direction ($y$ direction in the FIG. 1(c))

In COMSOL model, we use two-dimensional geometry, assuming continuity in $x$ direction putting the Floquet periodicity condition on the vertical boundaries. We neglect heat flow in $x$ and $z$ directions, and assume $\partial T/\partial x = \partial T/\partial z = 0$, the standard values for permalloy, ($c$ =495 J/kg/K, $\rho$=8700 kg/m$^3$, $k_T$=20 W/(m $*$K) [38, 39]), and glass ($c$=840 J/kg/K , $\rho$=2500 kg/m$^3$, $k_T$=1.1 W/(m

∗K) [40]). We assume the heat flow continuity at the permalloy–glass interface $k_i \frac{\partial T_i}{\partial y} = k_j \frac{\partial T_j}{\partial y}$, where *i, j* denotes permalloy or glass at the corresponding sides of the contact, and as a boundary conditions, use *∂T/∂y=0* at the upper and low boundaries of the COMSOL model (far enough from the structure). Heat loss via radiation [41] is estimated as $-k\frac{\partial T}{\partial y} = \in \sigma(T^4 - T_0^4)$, where $\in$ =0.3 is the emissivity of permalloy, and $\sigma$= 5.67 x $10^{-8} W m^{-2} K^{-4}$ is the Stefan-Boltzman constant.

The heat sources are described as

$$Q\ (y,\ t) = A * q\ (y) * I\ (t), \qquad (3)$$

where *q(y)* is loss calculated in the first step, FIG. 2 (c), *I(t)* is the temporal shape of the pulse, FIG. 2 (d), and *A* is the parameter estimated from our experimental conditions. We assume a gaussian profile for the pulse shape, $I\ (t) = exp\ (-t^2/\tau^2)$ with $\tau$ = 3 ns, which closely corresponds to the experimental shape of the pulse.

At typical experimental parameters (pulse energy of 0.1 mJ, absorption of ~ 35 % at 600 nm, spot diameter of 3 mm, and film thickness of 30 nm) using standard values of permalloy density and heat capacity, estimations show that the temperature of the film would rise by ~40 K in the absence of any heat exchange with surroundings.

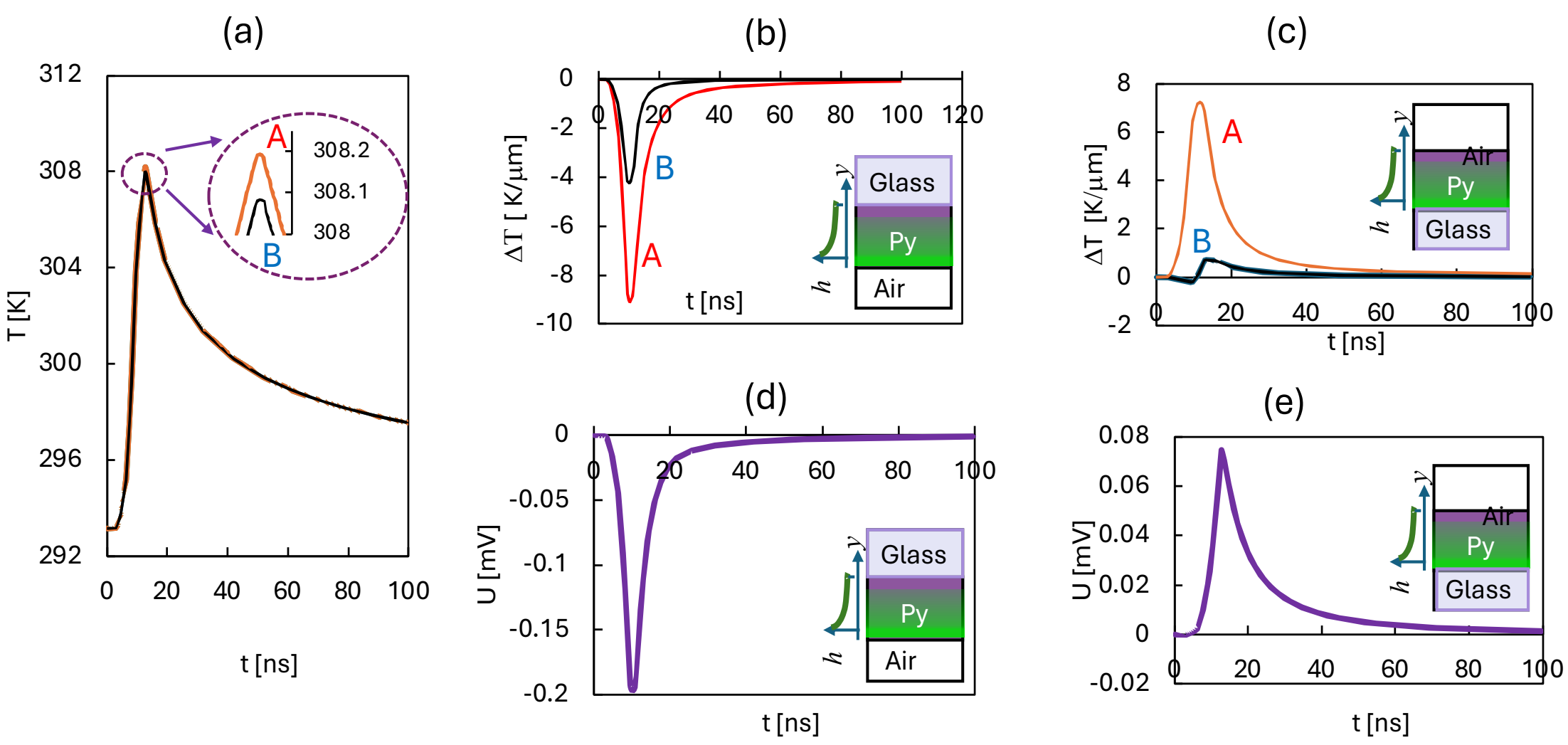

FIG. 3. Numerical simulations for temporal behavior of (a) temperature (at illumination from air), (b, c) temperature gradients in points A and B illuminated from (b) air and (c) glass and (d, e) ANE voltage at illumination from (d) air and (e) glass.

Account for thermal conduction to glass and radiation to air and glass results to a lower heating of the film, see FIG. 3(a). Since the heat sources are varied across the film, the temperature is slightly different across the film. To illustrate this, we show the temperatures and temperature gradients at the illuminated side of the film, Point A (5 nm from the bottom interface in FIG. 2 (a)) and the back side of the film, Point B (5 nm from the top interface). The maximum temperature difference is about 0.11 K, see inset in FIG.3(a). The temperature gradients in these two points are shown in FIG. 3(b) and (c) for two regimes of the illumination, from air and from glass.

Numerical simulations show that even if the losses decrease along *y* axis in both from air and from glass illumination conditions, the temperature gradients averaged across the film are determined by the presence of glass and show opposite signs for these two cases.

Using the Nernst coefficient $N\mu_0 M = -9$ nV/K [32], the estimated ANE photovoltages are shown in FIGs. 3(d, e). As one can see, the signal has the opposite polarities depending on the position of the glass substrate. The magnitude of the signal does not exceed 0.2 mV for typical experimental parameters.

## IV.EXPERIMENTAL RESULTS AND DISCUSSION

Experimental behavior of the photovoltages in permalloy structures is quite complicated and strongly depends on the structure geometry, wavelength, incidence angle, and magnetic field [23]. In this paper, we restrict ourselves only to the two experiments, which in our opinion, provide important information on the photovoltage behavior. In both experiments, the signals are recorded at two opposite magnetic fields of ±90 Oe, which are used to switch the magnetization of the sample between the directions in negative or positive *z*-axis direction. Since the coercive fields in permalloy can be as low as ~1 Oe, the application of the external magnetic field ensures the orientation of magnetization in a certain direction while in zero field, the domains might have a random magnetization depending on the magnetic history and/or small stray fields in the lab.

In the first experiment, we answer the following question: what affects the polarity of the magneto-dependent signal. Is it the gradient in the intensity of light across the film (as was suggested in [22]) or an asymmetry in the boundary conditions on the two interfaces (Py/air and Py/glass)?

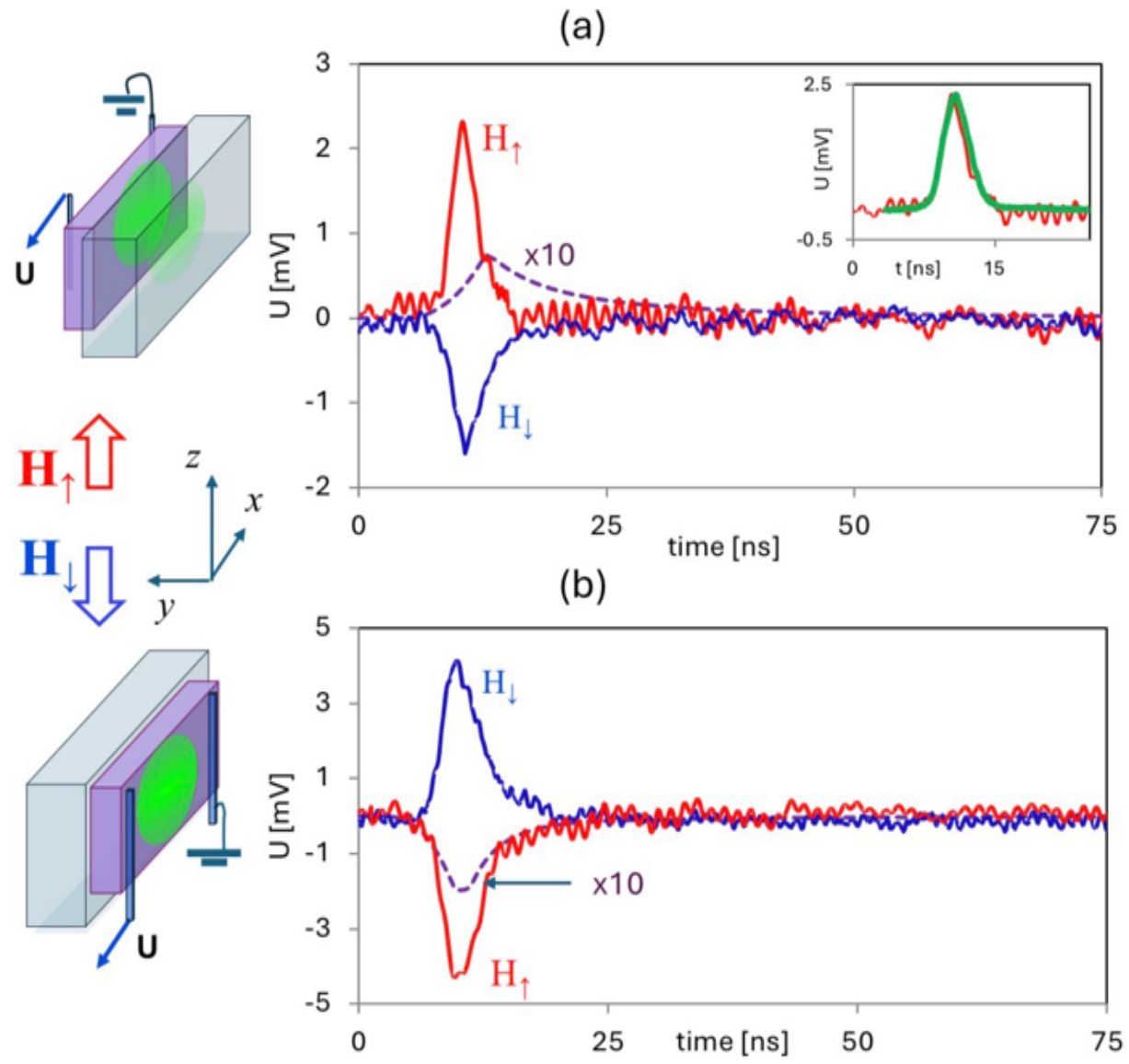

FIG. 4. Schematics of the illumination (left) and photoinduced electric signals (right) in Flat Py films when illuminated (a) from glass and (b) from air at upward (red) and downward (blue) directions of magnetic field. The dashed lines are predictions of ANE multiplied by 10. Inset represents comparison of the signal at upward field (red) with the shape of the laser pulse (green). The illumination wavelength λ = 600 nm, the pulse energy is about 0.1 mJ.

The flat film is illuminated at the normal incidence via glass (FIG. 4 (a)) or from air (FIG. 4 (b)). In both cases, the positions of electrical contacts, electrical connections and direction of light illumination are kept the same while the film is simply flipped by 180° around horizontal $x$ axis between two regimes of illumination. The signals are recorded in positive and negative fields in each case. The polarities are opposite in the opposite fields as expected and observed before; they are related to the film magnetization [22]. The most interesting finding is that the signal polarities depend on how the film is illuminated, through glass or through air. At the upward field, a signal is positive when the film is illuminated from glass and negative when illuminated from air. Note that in both cases, the gradient of light intensity has the same direction ($-y$) determined by the penetration of light in the permalloy film. If the signal is determined by this gradient, the polarity should not change. However, it switches to the opposite.

In principle, such behavior is in line with the predictions of the ANE effect, FIG. 3 (d, e). As discussed above, the temperature gradient is mostly dictated by the heat conduction to glass and has opposite directions for these two cases. In FIG 4, the predictions of ANE are shown with dashed traces. One can see that even if the opposite polarities of the signal for the opposite directions of illumination generally correspond to the predictions of ANE, the experimental magnitudes are much higher (more than an order of magnitude) than the theoretical estimations. This also corresponds to earlier estimations [22].

In addition, the kinetics of the experimental and predicted responses are different. The experimental kinetics mostly follows the temporal profile of the laser pulse (see inset in FIG. 4 (a)) while the numerical simulations predict longer kinetics for ANE voltage. Both growth and decay of the predicted signal are much slower than those observed in the experiment as most clearly seen from Fig. 4 (a). We should note that some experimental curves have a long "tail" which can be related to ANE effect, however, the main part of the signal requires a different mechanism for proper description.

The photovoltages taken at the normal incidence at upward and downward fields are of the opposite polarity and have similar magnitudes. However, at oblique incidence, the situation changes. In flat films, the angular dependence is weak [23], while in gratings, the signals change dramatically with varying the incidence angle. In the second experiment illustrated in FIG. 5(a, b), photovoltages in Py/Blu-ray sample are recorded at two incidence angles $\theta = 20°$ and -20., which correspond to the previously established plasmon resonance conditions [31].

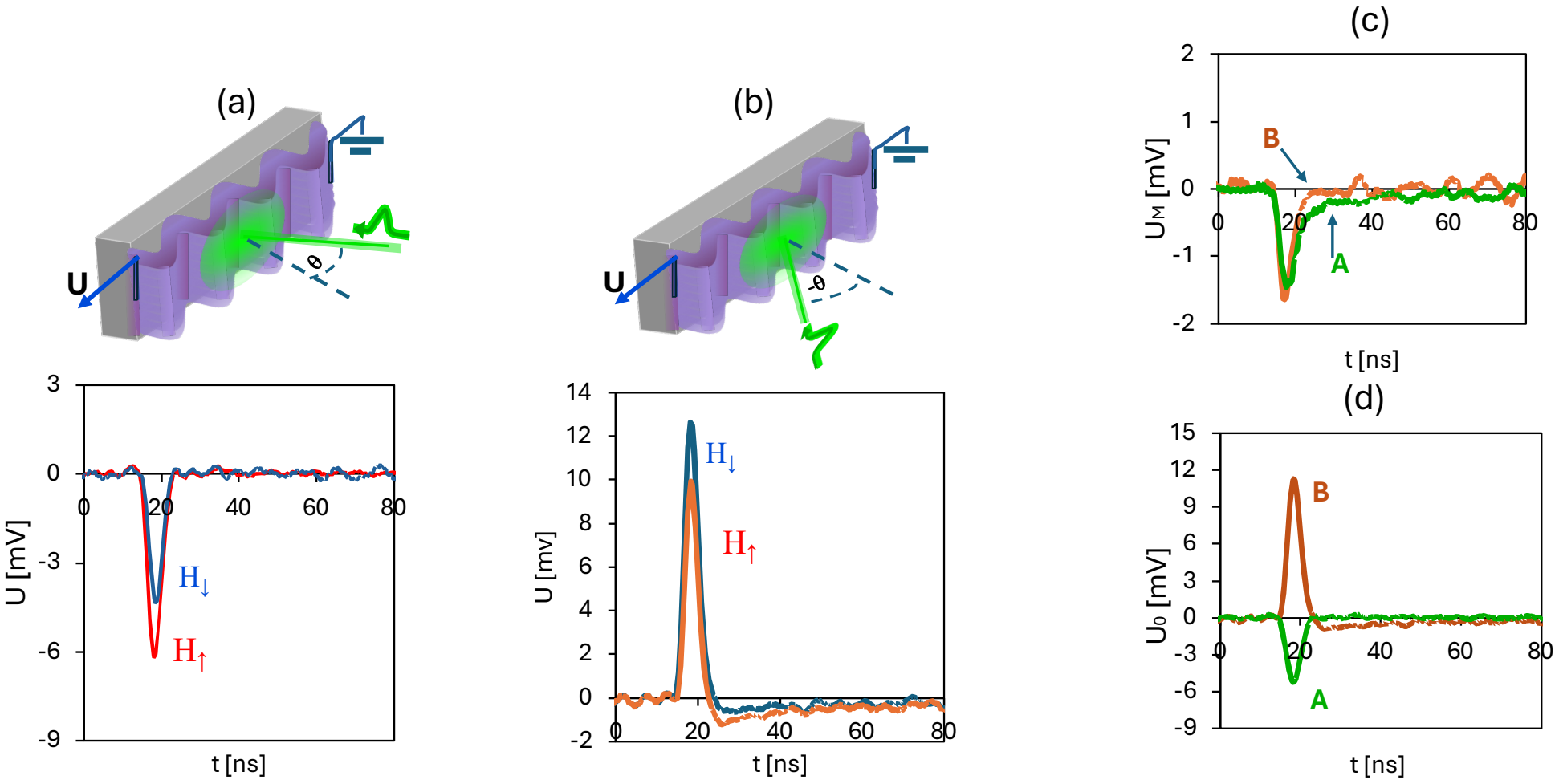

FIG. 5. (a) and (b) The schematics (top) and signals (bottom) in BluRay-based sample at upward (red) and downward (blue) directions of magnetic field at two opposite incidence angles; λ = 450 nm; (c) $U_M$ and (d) $U_0$ estimated from Eq.4 from Fig (a) (Trace A) and Fig (b) (Trace B) correspondingly.

As one can see, at 20°, signals $U\downarrow$ and $U\uparrow$ measured at respectively downward and upward fields are both negative (with different magnitudes) while at -20° they are both positive. Following [23], consider the photoinduced signal as a sum of two components, $U_M$ and $U_0$. The magneto-dependent part, $U_M$, changes the polarity depending on the field direction, and $U_0$ does not depend on magnetic field. $U_M$ and $U_0$ estimated from the experimental kinetics as

$$U_M = (U\uparrow - U\downarrow)/2 \text{ and } U_0 = (U\downarrow + U\uparrow)/2, \quad (4)$$

are shown in FIG. 5 (c, d).

The behavior of the magnetic component in Py/Blu-ray, $U_M$ is similar to the photovoltage observed in the flat film at the normal incidence. The non-magnetic $U_0$ is zero at the normal incidence, appears at the oblique incidence and has the opposite signs at the positive or negative angles. $U_0$ is of the same behavior and likely of the same origin as "plasmon drag" photovoltages observed in plasmonic metals, where it is greatly enhanced in nanostructured systems and strongly depend on nanoscale geometry maximizing at plasmon resonance conditions [3,10]. In permalloy, $U_0$ is very weak in flat films and significant in gratings where it can show a broad non monotonous dependence on angle [23]. The incidence angle of 20° used in our experiment corresponds to the plasmon resonance, however in permalloy the resonances are weak and broad due to high loss [31]. The kinetics of $U_M$, FIG. 5 (c), in gratings shows the presence of a small component with a longer decay in similarity to the kinetics in FIG. 4. $U_0$ rather follows the temporal behavior of the pulse, however the trace "B" has a small contribution with a long decay and of the negative polarity.

Thus, the main conclusions from this and previous studies [22,23] in permalloy thin films are the following:

- Photoinduced electric effects in permalloy have magnetically dependent and magnetically independent components, which both are very sensitive to the nanoscale geometry and illumination conditions.
- The polarity of the magnetically dependent component is determined by magnetization and depends on asymmetry in the film interfaces (shows opposite signs at glass/Py/air or air/Py/glass illumination regimes).
- The magnetically independent component is greatly enhanced when the sample is nanostructured and depends on the light propagation direction in similarity to the plasmon drag effect observed in silver, gold and other plasmonic nonmagnetic materials.
- Theoretical estimations of ANE photovoltages predict qualitatively correct behavior, however predicted magnitudes are much weaker, and their kinetics is slower than those observed in the experiment. Note that at the same time, the plasmonic pressure model [20, 21] for the plasmon drag effect correctly predicts its angular behavior and fails to describe the effect quantitively (orders of magnitude weaker than the experiments [16]).

Let us now discuss a possible scenario for both components, taking into account that the predominant part of the photovoltage is observed during laser pulse well following its temporal profile. Small contributions to the signals with longer decay can be partially or fully related to thermoelectric effects while the origin of the main part is associated with excited electrons which are not in the equilibrium with lattice.

In ferromagnetic material the positions of bands are different for majority and minority spins, see [42,43] for the detailed band structure of permalloy. The probabilities of inter-band transitions [44] and photoelectron excitation and emission [41] depend on spin orientation. The laser pulse creates a disturbance in spin polarization leading to spin currents and charge currents via Inverse Spin Hall Effect. Generation of spin currents is expected to be strongly enhanced at plasmon resonance [45].

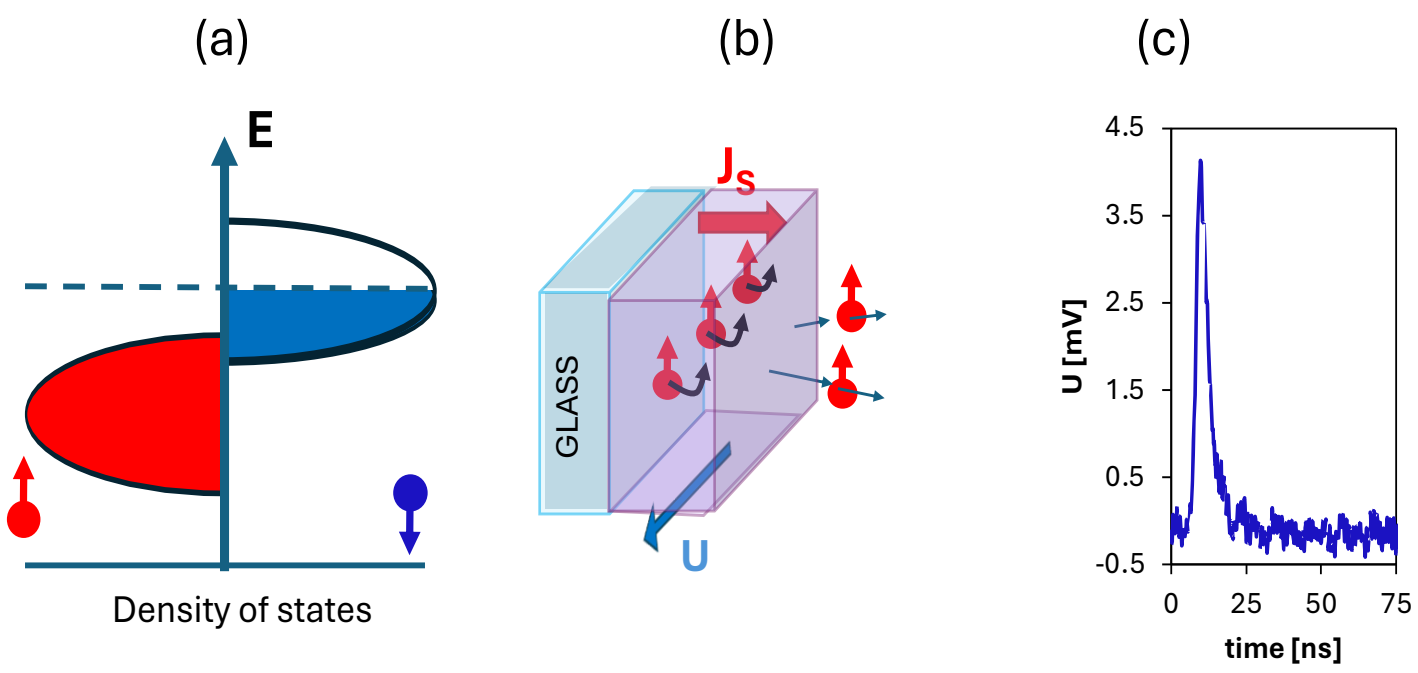

FIG. 6. (a) Simplified band diagram; (b) Schematics of spin currents and ISHE voltage due to photoelectron emission; (c) experimental signal in flat film in corresponding geometry.

To explain the opposite polarities of the magnetically dependent component observed at illumination from glass or from air, let us take into account emission of hot electrons which can be excited by strong laser light. In visible range, the majority of emitted electrons are majority spins [43]. It can be expected that electrons are emitted predominantly to air and not into glass (where they would form a charged layer preventing the further emission). When the film is illuminated from air, FIG. 6 (b), the emission of majority spins leads to the spin current directed to Py/air interface, and a positive voltage according to ISHE. When the sample is flipped, having Py/air interface on the back side and glass on the front side, the spin current and ISHE voltage switch polarities. This corresponds to the experimental behavior of the magnetically dependent component, FIG. 4.

Although ISHE has been experimentally demonstrated in permalloy [34,35], the reported voltages under conventional spin-pumping conditions are typically in the microvolt range, much lower than several millivolts observed in our experiments. However, in contrast to dc experiments, in our case, we have highly non-equilibrium dynamics of hot electrons excited by strong laser pulses. Laser-induced non-equilibrium spin transport and ultra-fast demagnetization processes are reported in magnetic or magnetic-normal metal structures under femtosecond laser pulse illumination [50-53].While the relaxation times for hot electrons are in picosecond or sub-picosecond scale [50], the hot electron transport can have ballistic or superdiffusive regime [51] featuring a very long propagation length, $l_s$ and leading to significantly enhanced transient spin currents. (As an example, values of $l_s$ = 100 nm are reported in gold [54]). Further experimental studies with short laser pulses and various film thickness would provide more information on parameters of non-equilibrium spin transport in permalloy and its possible contribution to the observed effects.

To explain the magnetically independent "plasmon drag" component, let us take into account the Doppler effect following the model [27,28]. For electrons with velocities antiparallel to the projection of photon k-vector onto film plane (or to plasmon k-vector), the frequency is higher than for those moving parallel to k-vector. In plasmonic structures, this can lead to different probabilities of generation and multiplication of hot carriers with the opposite velocities through Landau damping. For the visible range, one can expect the excitation and emission of hot electrons to be more efficient for electrons with the velocities against k-vector, resulting in changes in the distribution of electron velocities and a net electron drift parallel to the photon k-vector. Nanostructurization, impurities, and roughness leading to localized plasmon excitations play significant role in hot electron emission [46-49]. This corresponds to experimental findings in significant enhancement of plasmon drag effect in rough films and nanostructures [3,10, 23].

In conclusion, kinetics of transverse photovoltages excited by nanosecond laser pulses in permalloy thin films are studied at different photoexcitation configurations. The signals are analyzed as sum of magnetically dependent and magnetically independent components. Predominant parts of both components follow the temporal profile of the laser pulse while simulations of thermoelectric ANE effect predict the voltages of much lower magnitude and slower kinetics. Non-equilibrium spin transport and photoemission of hot electrons is suggested as a possible main mechanism for photovoltage generation induced by short laser pulses.

## ACKNOWLEDGMENTS

This work was supported by the National Science Foundation under Grant NSF EIR 2301350 and Grant NSF CREST 2112595, and the Department of Energy under Grant DOE NNSA DE-NA0004007. We also acknowledge the cooperation from LLNL to help us deposit permalloy films using their Sputtering facility.

## DATA AVAILABILITY

The experimental and numerical model data are available from the authors upon reasonable request.